\documentclass[
    ,final            
  ]
  {aipproc}

\layoutstyle{6x9}

\usepackage{psfig}
\usepackage{amsmath,bbm,amssymb}

\def\gtap{\mathrel{\hbox{\rlap{\lower.55ex \hbox {$\sim$}}
                   \kern-.3em \raise.4ex \hbox{$>$}}}}
\def\ltap{\mathrel{\hbox{\rlap{\lower.55ex \hbox {$\sim$}}
                 \kern-.3em \raise.4ex \hbox{$<$}}}}

\begin{document}

\title{X-ray sources in globular clusters of other galaxies}


\author{Walter H.G. Lewin}{
  address={Massachusetts Institute of Technology, Physics Department
Center for Space Research, MA 02139, USA}
}
\author{Frank Verbunt}{
  address={Astronomical Institute, Postbox 80.000, 3508 TA Utrecht,
the Netherlands}
}

\begin{abstract}
A large number of X-ray sources in globular clusters of galaxies other
than the Milky Way has been found with Chandra.  We discuss three
issues relating to these sources. The X-ray luminosity function (XLF)
of the sources in globular clusters of M31 is marginally compatible
with the XLF of globular clusters of the Milky Way. The 
individual XLFs of a dozen elliptical galaxies, {\it after
correction for incompleteness}, are compatible with one another
and show no break; however, the XLF found by adding the individual XLFs of
elliptical galaxies has a break at $L_x\simeq5\times10^{38}$ erg\,s$^{-1}$.
For the moment there is no evidence for a difference between the
XLFs of sources inside and outside globular clusters of elliptical
galaxies. It is not (yet?) possible to decide which fraction
of low-mass X-ray binaries in elliptical galaxies outside globular
clusters have formed inside globular clusters.
\end{abstract}

\maketitle


\section{Introduction}

Observations with Chandra of an increasing number of nearby galaxies
are revealing a sizable population of bright X-ray sources, many of
which are in globular clusters. In elliptical galaxies, as many as
half of all bright X-ray sources are in globular clusters; in spiral
galaxies a smaller fraction of the bright sources is in globular
clusters. The absence of recent star formation in elliptical galaxies
implies that the X-ray sources in them are mainly low-mass X-ray
binaries.  Some questions that have arisen from early research are
whether globular clusters do contain black holes? and whether the
sources in elliptical galaxies outside globular clusters originate in
globular clusters?

Elliptical galaxies are a good target for the study of X-ray sources
in globular clusters, because many have 5000 to 10\,000 globular
clusters. The central galaxy of the Virgo cluster, M\,87, has 13500
globular clusters!
The X-ray sources detected so far in the faraway galaxies are of
necessity {\em very} bright, typically $L_x>5\times10^{37}$ erg\,s$^{-1}$.
Early studies indicate that about 4\%\ of the globular clusters
contains an X-ray source above this limit, roughly corresponding to
1 source per $5\times10^{6}L_{\odot,I}$ (Sarazin et al.\ 2003, Kundu
et al.\ 2003). 
\nocite{kmzp03}\nocite{ski+03}

We have given an overview of the early papers on this topic in
Verbunt \&\ Lewin (2005).
Comparison between studies is difficult, as they
have different detection limits and compute X-ray luminosities in
different energy bands based on different assumed spectra.
The study of Kim \&\ Fabbiano (2004) remedies this, and provides a
first consistent comparison between a fair number (14)
of different elliptical galaxies;
we discuss this important study in Sect.\,3.
In Sect.\,2 we compare the X-ray luminosities of globular clusters 
in our galaxy and in the Andromea nebula. In Sect.\,4 we discuss
the question whether sources outside globular clusters can have formed
inside them.
\nocite{vl05}\nocite{kf04}

\section{Globular clusters in the Milky Way and in M31: different
or the same?}

\begin{figure}
\centerline{\psfig{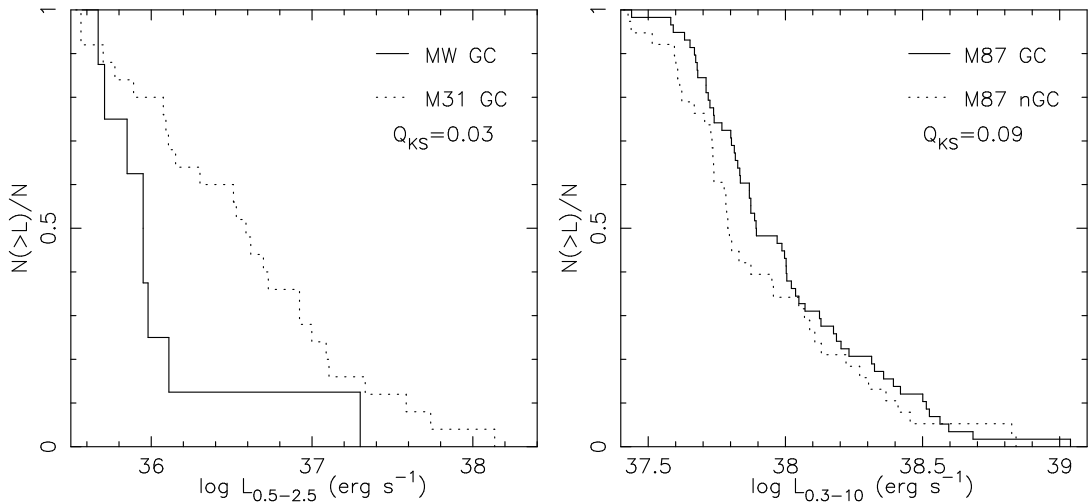}}

\caption{\it Left: Comparison of the normalized cumulative X-ray luminosity 
functions (at $L_x>10^{35.5}$ erg\,s$^{-1}$) of globular clusters
of our Milky Way and of M\,31.
The Chandra luminosities given by Di Stefano et al.\ (2002) were
multiplied by 0.46 to convert them to the energy range of ROSAT data
from Verbunt et al.\ (1995).  After Verbunt \&\ Lewin (2005).
Right: Comparison of the normalized cumulative X-ray luminosity 
functions (at $L_x>10^{37.4}$ erg\,s$^{-1}$) of M87 sources in
globular clusters and outside them. After Jord\'an et al.\ (2004). 
The Kolmogorov-Smirnov probability that the normalized distributions
are the same is 0.03 for the Milky Way vs.\ M\,31, and 0.09 for
in vs.\ out globular clusters of M\,87.
\label{lewinfa}}
\end{figure}

The Andromeda Nebula, a.k.a. M\,31, was found with Einstein to have
rather more bright X-ray sources in globular clusters than our Milky
Way (Van Speybroeck et al.\ 1979). Since this discovery it has been
debated whether this simply reflects the larger number of globular
clusters of M\,31, or an X-ray luminosity function with a higher
fraction of bright sources in M\,31 (e.g. Verbunt et al.\ 1984).  The
debate continued in the ROSAT and Chandra era (Magnier et al.\ 1992,
Supper et al.\ 1997, Di Stefano et al.\ 2002).  In
Figure\,\ref{lewinfa} we show the cumulative X-ray luminosity
functions for globular clusters of our Milky Way and of M\,31.  A
2-sided Kolmogorov-Smirnov test gives a probability of 3\%\ that both
distributions are the same. Thus, the evidence that the luminosity
functions are intrinsically different is at the 2-sigma level:
suggestive, but not conclusive.
\nocite{sef+79}\nocite{vpe84}\nocite{mlp+92}\nocite{shp+97}\nocite{skg+02}
\nocite{vbhj95}

The X-ray luminosity functions of the globular clusters in 
M\,31, and the bulge in M\,31 (Kong et al.\ 2002, 2003)
are both roughly compatible with the X-ray luminosity function of
elliptical galaxies (see next Section), an indication that the X-ray
sources in all these old stellar populations are similar. (With a
population of eight persistent sources and 5 transients, the globular
cluster system of our galaxy has too few bright X-ray sources to 
constrain the slope of luminosity function well.)
\nocite{ksgg03}\nocite{kgp+02}

\section{X-ray luminosity functions of globular cluster systems of elliptical
galaxies}

Kim \&\ Fabbiano (2003) investigated the incompleteness at the
low-luminosity end of the X-ray luminosity function of the elliptical
galaxy NGC\,1316, and showed that the apparent break in the XLF
disappears when appropriate corrections are made. They then
investigated XLFs of 13 more elliptical galaxies, and showed that none
of these shows a significant break after correction (Kim \&\ Fabbiano
2004, see also Gilfanov 2004). The elliptical galaxies comprise 7
galaxies of the Virgo cluster (at 17 Mpc), 2 of the Fornax cluster
(19.9 Mpc), and 5 others (between 11 and 29 Mpc).  The effects leading
to incomplete detection efficiency at low X-ray luminosities are
\begin{itemize}
\item the presence of diffuse emission of the hot interstellar medium 
in E and S0 galaxies
\item the Eddington bias, enhancing the number of sources near the
detection threshold
\item source confusion
\item the larger point-spread function near the detector edge
\end{itemize}
\nocite{kf03}\nocite{gil04}

The X-ray luminosities are determined in the energy range
0.3-8.0\,keV; and sources are counted outside the innermost region of
each galaxy ($r>20''$) but within the 25 mag isophote. Corrections
for the source numbers and total X-ray luminosity of the galaxy are
made on the assumption that they scale as the optical luminosity,
i.e.\ with $r^{1/4}$: this affects the normalization, but not the
slope of the luminosity function. It should be noted that Kim \&\
Fabbiano (2003, 2004) do not discriminate between sources within and
outside globular clusters, and give the XLFs of all sources related to
elliptical galaxies. They find that the XLF of every individual
elliptical galaxy is compatible with a power law $N(L_x)\propto
{L_x}^{-\beta}$, with $\beta=2.0\pm0.2$.  The variation between the
$\beta$-values of different ellipticals can be ascribed to
small-number statistics. In studies of individual elliptical galaxies
it is found that the XLF for sources in globular clusters is the same
as the XLF for sources outside globular clusters (Maccarone et al.\ 2003,
Sarazin et al.\ 2003, Jord\'an et al.\ 2004, see Fig.\,\ref{lewinfa}).
Thus, we may take the overall XLF as a proxy for the XLF of the
globular cluster sources.
\nocite{mkz03} 

\begin{figure}
\centerline{\psfig{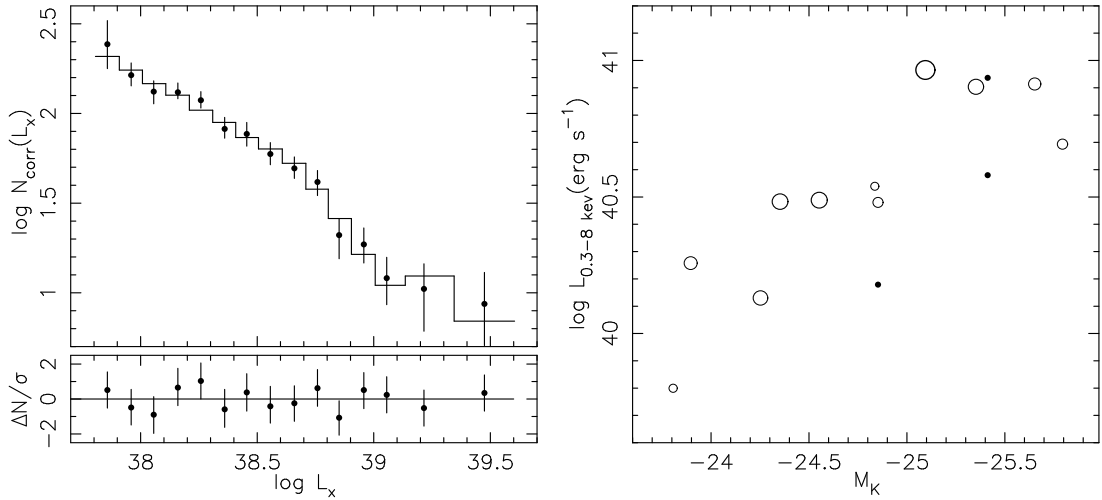} {\hfil}}
 
\caption{\it Left: total corrected number of sources in 14 elliptical
galaxies as a function of X-ray luminosity in the 0.3-8.0\,keV range.
Right: corrected total X-ray luminosity of elliptical galaxies as a
function of their absolute K-band magnitude $M_K$. The total X-ray
luminosity is computed from the corrected X-ray luminosity function,
extrapolated downwards to $L_x=10^{37}$ erg\,s$^{-1}$. The size of the
symbols $\circ$ scales with the number of globular clusters
per unit luminosity of the galaxy $S_N$. $\bullet$ indicates a galaxy for
which $S_N$ is not known. (After Kim \&\ Fabbiano 2004.) \label{lewinfb}
}
\end{figure}

Since the individual X-ray luminosity functions appear to be the same,
they can be added. The added XLF, at luminosities in the 0.3-8.0 keV
band $L_x>6\times 10^{37}$ erg\,s$^{-1}$, of all 14 elliptical  galaxies
is marginally compatible with a single power law with $\beta=2.1\pm0.1$.
A broken power law gives a somewhat better fit, and is shown in 
Figure\,\ref{lewinfb}. The best-fit break luminosity is
(5.6$\pm$1.6)$\times 10^{38}$ erg\,s$^{-1}$, and the values for
$\beta$ at lower and higher luminosities are 1.8$\pm$0.2 and
2.8$\pm$0.6, respectively.
Remarkably, the break is at about the flux where the luminosity function
of AGNs and quasars also shows a break. Whereas background sources typically
comprise about 5\%\ of the sources within the 25\,mag isophote, they
are unlikely to cause the break in the added  XLF of the elliptical
galaxies, which has been corrected for the background sources
(Kim \&\ Fabbiano 2004).

The break is at a luminosity which is about twice the Eddington limit for
hydrogen-rich material, and comparable to the Eddington limit for
hydrogen-poor material (see also Kuulkers et al.\ 2003).  From their
high X-ray luminosity and soft X-ray spectrum, it has been argued that
the sources above the break are accreting black holes (e.g.\ Angelini
et al.\ 2001). As regards the sources below the break, Bildsten \&
Deloye (2004) note that the assumption that most very luminous X-ray
sources are ultracompact binaries, in which the donor star is a
hydrogen-depleted degenerate star, explains the location of the break
at the Eddington limit of hydrogen-poor material, the high X-ray
luminosities, the source incidence (or birth rate), and the X-ray
luminosity function. The model does not explain why sources outside
globular clusters, which are less likely to be ultracompact binaries
(if our own Milky Way is a guide), have the same X-ray luminosity
function. This brings us to the question whether sources outside
globular clusters could have formed inside them.
\nocite{khz+03}\nocite{alm01}\nocite{bd04}

\section{Outside and inside globular clusters}

In elliptical galaxies it is found that the spatial distribution of X-ray
sources outside globular clusters is similar to the spatial distribution
of globular cluster X-ray sources (e.g. Jord\'an et al.\ 2004 for M\,87).
Similarly, the X-ray luminosity functions of globular cluster
sources and other sources are very similar (Fig.\,\ref{lewinfa}).
This has led to the revival of a suggestion by Grindlay \&\ Hertz (1985)
that all low-mass X-ray binaries, including those now outside globular
clusters, were formed inside globular clusters (White et al.\ 2002).
\nocite{gh85}\nocite{wsk02}

In the Milky Way and in M\,31 there are about 10 bright low-mass X-ray
binaries in the disk for each one in a globular cluster. In elliptical
galaxies, there is about 1 source outside globular clusters for each
one in them. This suggests that the majority of low-mass X-ray
binaries in the disk of the Milky Way , M\,31 and by generalization in
spiral galaxies are formed in the disk, and not in globular clusters
(Verbunt \&\ Lewin 2005). It should be noted that we cannot compare fractions
inside and outside globular clusters for the Milky Way and M\,31 
on one hand and elliptical galaxies at the other hand {\em at
the same luminosities}, because there are very few X-ray sources
in the Milky Way and M\,31 above $L_x\sim 5\times10^{37}$ erg\,s$^{-1}$,
which is the lower limit of detectable sources in ellipticals
(see Fig.\,\ref{lewinfa}).

The total X-ray luminosity of low-mass X-ray sources in elliptical
galaxies scales with the luminosity (hence presumably stellar mass) of
the elliptical galaxy (Figure\,\ref{lewinfb}). This is expected for
sources formed outside globular clusters.  It is also expected if the
sources are mainly formed in globular clusters, because the number of
globular clusters is on average higher in large elliptical galaxies
than in small ones. Correlations of the total X-ray luminosity have
been made with the specific frequency $S_N$ of globular clusters
(i.e.\ the total number of globular clusters divided by the luminosity
of the galaxy). We refer to this here as the "global" specific
frequency which includes all globular clusters in the entire galaxy.
These correlations
greatly suffer from a serious problem related to our lack of knowledge
of the global $S_N$. Reliable values for the number of globular clusters, in
general, come from HST-WFPC2 with a very small (5.7 square arc minutes)
field of view. This provides only a reliable local value for $S_N$
but not the global value (see the discussion in Section
8.3 of Verbunt \& Lewin 2005).

Kim \&\ Fabbiano (2004) note that the correlation between the total
X-ray luminosity (of low-mass X-ray sources) and the infrared
luminosity of elliptical galaxies has a scatter which is larger than
the measurement accuracy. In Fig.\,\ref{lewinfb} we indicate the
elliptical galaxies with a large $S_N$ with a large symbol. We see
that the elliptical galaxies brighter in X-rays than the average
correlation indeed have a tendency to have a higher $S_N$. This
indicates that globular cluster sources contribute significantly to
the total X-ray luminosity of the galaxy; to argue that sources
outside globular clusters originate inside globular clusters, one
would have to show that the total X-ray luminosity (or better, number)
of the sources {\em outside} globular clusters separately scales with specific
frequency $S_N$.  In the sample of galaxies studied by Kim \&\
Fabbiano (2004) there is {\em no} correlation between $M_K$ and $S_N$
(see Fig.\,\ref{lewinfb}).

\section{Conclusions and Outlook}

\begin{figure}
\centerline{\psfig{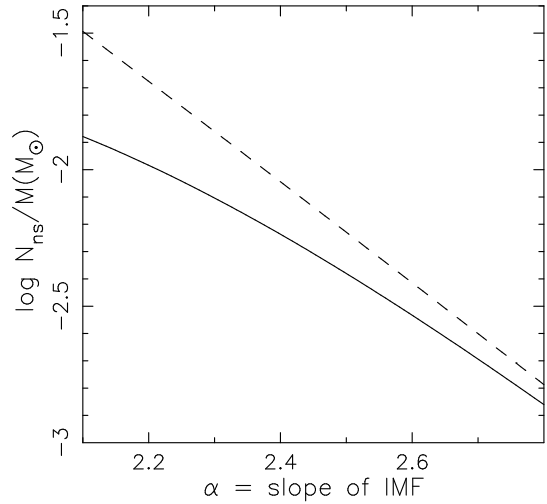} {\hfil}}
 
\caption{\it The number of neutron stars per solar mass formed
from an initial mass distribution $N(m)\propto m^{-\alpha}$ 
as a function of the index $\alpha$. The initial range of stellar
masses is taken to be from 0.08 to 80 $M_\odot$, and of neutron-star 
progenitors
from 8 to 80 $M_\odot$. The solid curve is normalized on the
{\em initial} mass of the cluster, the dashed curve on the {\em current}
luminous mass, assuming that stars from 0.8 to 8 $M_\odot$
have evolved into white dwarfs. \label{lewinfc}
}
\end{figure}

The conclusions drawn above, i.e.\ that the X-ray luminosity
function of the X-ray sources in old populations is universal
and has a break at $L_x\simeq5\times10^{38}$ erg\,s$^{-1}$, are based
on a study which does not discriminate between sources inside and
outside globular clusters. With the increasingly large sample
of sources it is possible and necessary to repeat this study
for the globular cluster sources separately. This may also help
in adressing the question which fraction of the sources outside
globular clusters were actually formed inside them.
The location of the break suggests that the sources above the
break are accreting black holes. The sources below the break may be
dominated by ultracompact sources (Bildsten \&\ Deloye 2004).

There are two important areas of further study not
discussed above. The first is the question why clusters with a high
metallicity have a higher probability of containing an X-ray source,
even if the number of collisions in them is the same. An interesting
calculation by Jord\'an et al.\ (2004) shows that a moderate
dependence on metallicity of the slope of the Initial Mass Function
leads to a difference in the numbers of neutron stars in metal-rich
and metal-poor clusters (per unit mass) which is large enough to
explain the observed preference for high-metallicity clusters
(see Fig.\,\ref{lewinfc}). Briefly, if we write the number of stars
at mass $m\equiv M/M_\odot$ as $N(m)=Km^{-\alpha}$, we have for the number
of stars with initial mass between $m_1$ and $m_2$ (that have 
evolved into neutron stars) per unit mass between $m_a$ and $m_b$:
\begin{equation}
{N(m_1,m_2)\over M(m_a,m_b) (M_\odot)} = {2-\alpha\over1-\alpha} \quad
{m_1^{1-\alpha}-m_2^{1-\alpha}\over m_a^{2-\alpha}-m_b^{2-\alpha}}
\label{lewineq:ns}\end{equation}
Jord\'an et al.\ (2004) discuss the number of neutron stas per unit of
{\em initial} mass of the cluster, i.e.\ they take $m_b=m_2$.  This
ratio, for values $m_a=0.08$, $m_b=m_2=80$ and $m_1=8$, is shown as a
solid line in Fig.\,\ref{lewinfc} as a function of $\alpha$. If
$\alpha$ is lower at higher metallicity, i.e.\ metal-rich clusters
have a shallower initial mass function, then metal-rich clusters have a
relatively higher number of neutron stars.  We may add that this
effect is even stronger if we scale not on the initial mass, but on
the current luminous mass, i.e.\ the stars that determine the observed
$K$ magnitude. Eq.\,\ref{lewineq:ns} for $m_2=80$ as before, but with 
$m_b=0.8$,  is shown in Fig.\,\ref{lewinfc} as a dashed line. (Stars in
the range $0.8<m<8$ have evolved into white dwarfs.)  This deserves
further study, in particular whether the required dependence of the
slope of the IMF on metallicity can be more firmly established.  

The second research area is the indication that the probability for a
globular cluster to contain an X-ray source increases slower with
density than the number of collisions occurring in the cluster
(Jord\'an et al.\ 2004). The models for formation and evolution of
X-ray sources in globular clusters must address these questions.





\bibliographystyle{aipprocl} 


\end{document}